\documentclass[11pt, a4paper]{article}
\usepackage{moriond,epsfig}

\def\Mbc{M_{\rm bc}}
\def\de{\Delta E}

\def\mco{\multicolumn}

\def\be{\begin{equation}}
\def\ee{\end{equation}}
\def\bea{\begin{eqnarray}}
\def\eea{\end{eqnarray}}

\begin{document}
\vspace*{4cm}
\title{Hadronic B decays at BELLE and BABAR}

\author{ S.-W. Lin}
\address{Physics Department, National Taiwan University, Taipei 106, Taiwan}
\maketitle\abstracts{There are several exciting results for the hadronic $B$ 
decays of Belle and BaBar recently. My report focuses on the ratios of branching fractions and $CP$
asymmetry for $B \to hh$ decays at Belle and BaBar where $h$ denotes $\pi$ or 
$K$. The observations of $B^+ \to K^+ \overline{K}{}^0$, $B^0 \to K^0 \overline{K}{}^0$ are published both by Belle and BaBar and shown here. We also report
the observation of $B^+ \to \rho^+ K^0$ and search of $B^+ \to \overline{K}{}^{*0} K^+$, $B^0 \to \rho^0 \rho^0$, $B^0 \to a_0^- \pi^+$ and $B^0 \to a_0^- K^+$
at BaBar.
Finally, we will show the results of amplitude analysis of the decays 
$B^0 \to \phi K^*_2(1430)^0$, $\phi K^*(892)^0$ and $\phi (K\pi)^0_{\rm S-wave}$ at BaBar.} %

\section{Introduction}
In general, the branching 
fractions with the SM predictions suffer from large hadronic uncertainties 
within the current theoretical framework and many of the uncertainties
cancel out in ratios of branching fractions. We report the ratios $R_{c,n}$
of the charged and neutral $B \to K \pi$ branching fractions and direct $CP$
asymmetry of $B \to K \pi$. Theoretical predictions with different approaches suggest 
that $A_{CP}(K^+ \pi^-)$ could be either positive or negative \cite{acp3}. Although 
there are large uncertainties related to hadronic effects in the theoretical 
predictions, results for $A_{CP}(K^+ \pi^-)$ and $A_{CP}(K^+ \pi^0)$ are expected to 
have the same sign and be comparable in magnitude \cite{acp3}.
The observation of $B^0 \to K^0 \overline{K}{}^0$
and $B^+ \to \overline{K}{}^0 K^+$ expected to be dominated by the loop-induced
$b\to d\overline ss$ process (called a $b\to d$ penguins) are published by Belle
and BaBar recently. BaBar also observed $B^+ \to \rho^+ K^0$ decay which is 
expected to be a pure penguin decays and the result helps to separate the
contribution of tree and penguin amplitudes in other channels. For the
amplitude analysis, we will show the results of $B^0 \to \rho^0 \rho^0$ and 
$B^0 \to \phi K^*_2(1430)^0$, $\phi K^*(892)^0$ and $\phi (K\pi)^0_{S-{\rm wave}}$. The nature of the $a_0$ meson is still not well understand and the
branching fractions of $B^0 \to a_0^- \pi^+$, $B^0 \to a_0^- K^+$,
$B^0 \to a_0(1450)^- \pi^+$ and $B^0 \to a_0(1450)^- K^+$ will provide some
information about the nature of $a_0$.

\section{$B \to K \pi$, $\pi \pi$ and $K K$ }
The new measurements of the branching fractions for 
$B \to K^+\pi^-$, $K^+ \pi^0$, $K^0 \pi^0$, $\pi^+ \pi^-$ and  
$\pi^+ \pi^0$ at Belle \cite{br1} and BaBar \cite{br2,br3} are shown in Table \ref{br}. The effect 
of final-state radiation is considered in branching fraction measurement 
now. The statistical errors on the branching fraction for all decay
modes are reduced. The ratio $R_c$ ($R_n$) obtained by Belle and BaBar's experimental results are $1.08 \pm 0.06 \pm 0.08$ ($1.08 \pm 0.08 ^{+0.09}_{-0.08}$) and $1.11 \pm 0.07$ ($0.94 \pm 0.07$), respectively. The
current $R_c$ and $R_n$ haved moved quite a bit towards the SM predictions and reduce the ``$B \to K \pi$
puzzle''. The observation of $B^0 \to \overline{K}{}^0 K^0$ and $B^+ \to \overline{K}{}^0 K^+$ by Belle \cite{kk1} and BaBar \cite{kk2} are shown in Table \ref{br} and the results agree with some
theoretical predictions \cite{plb253,plb341,pqcd_kk,ben,fleischer}.

\begin{figure}
\centering
\epsfig{figure=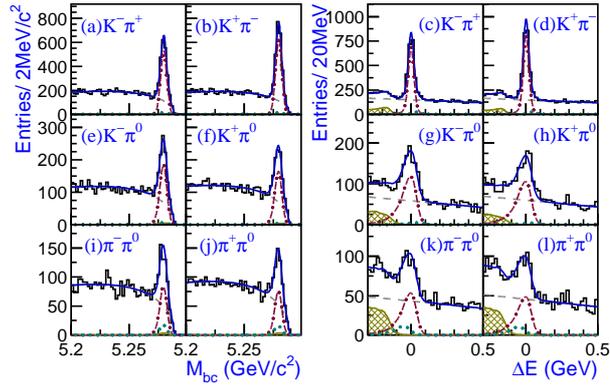,height=2.5in}
\caption{$M_{\rm bc}$ (left) and $\Delta E$ (right) distributions for
$B^0\to K^+\pi^-$, $B^+ \to K^+ \pi^0$ and $B^+ \to \pi^+ \pi^0$ candidates at Belle.
 The histograms show 
the data, while the curves represent the various components from
the fit: signal (dot-dashed), continuum (dashed), charmless $B$ decays
(hatched), background from  mis-identification (dotted),
and sum of all components (solid). The $\Mbc$ and $\de$ projections of the fits
 are for events that have $|\de|< 0.06$ GeV (left) and 
$5.271$ GeV/$c^2 < {M_{\rm bc}} <5.289$ GeV/$c^2$ (right). (A looser 
requirement, $-0.14$ GeV $<\de<0.06$ GeV, is used for the modes with a $\pi^0$ 
meson in the final state.)}
\label{belle:kpi}
\end{figure}

\begin{figure}
\centering
\epsfig{figure=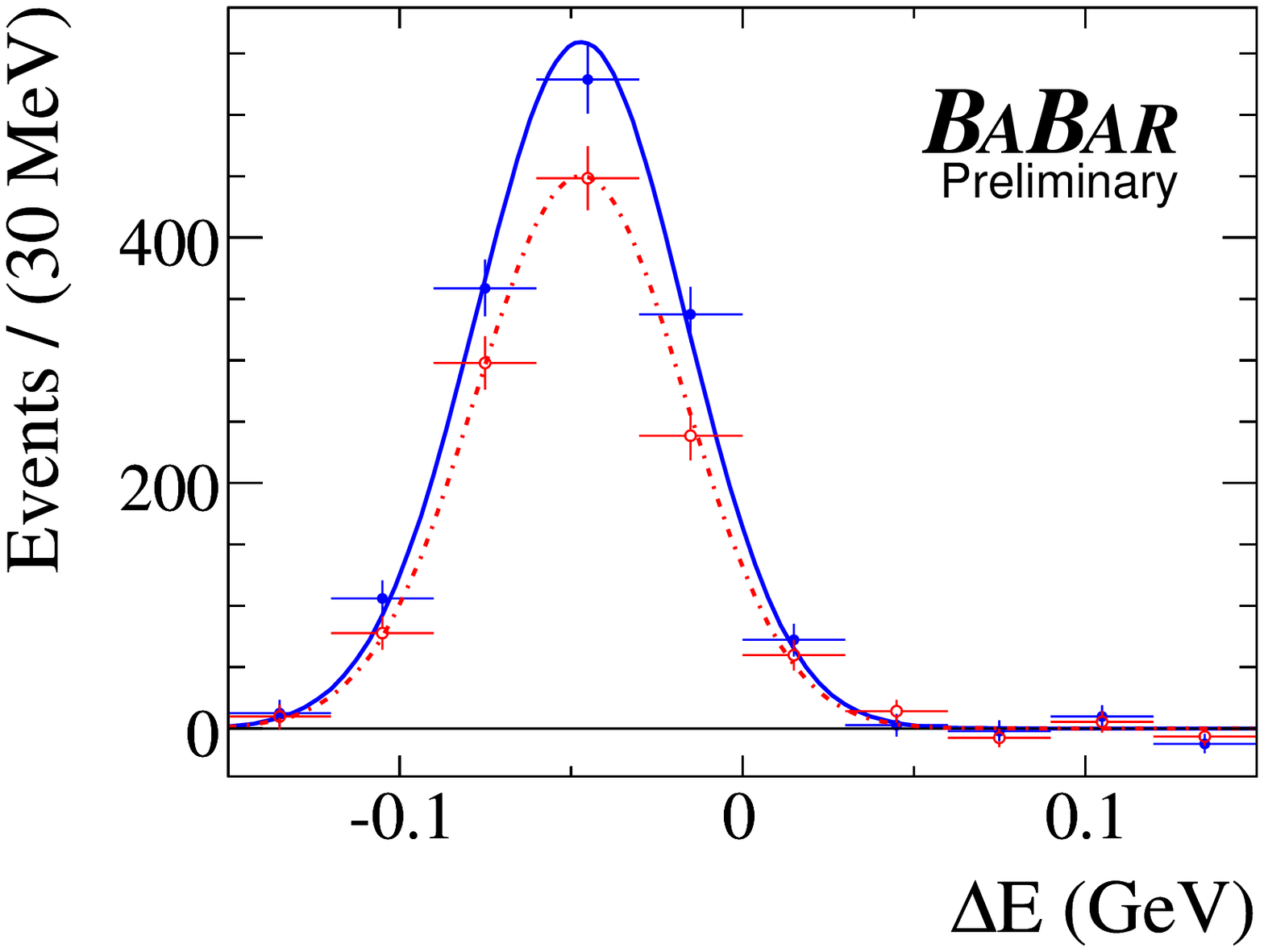,height=1.3in}
\epsfig{figure=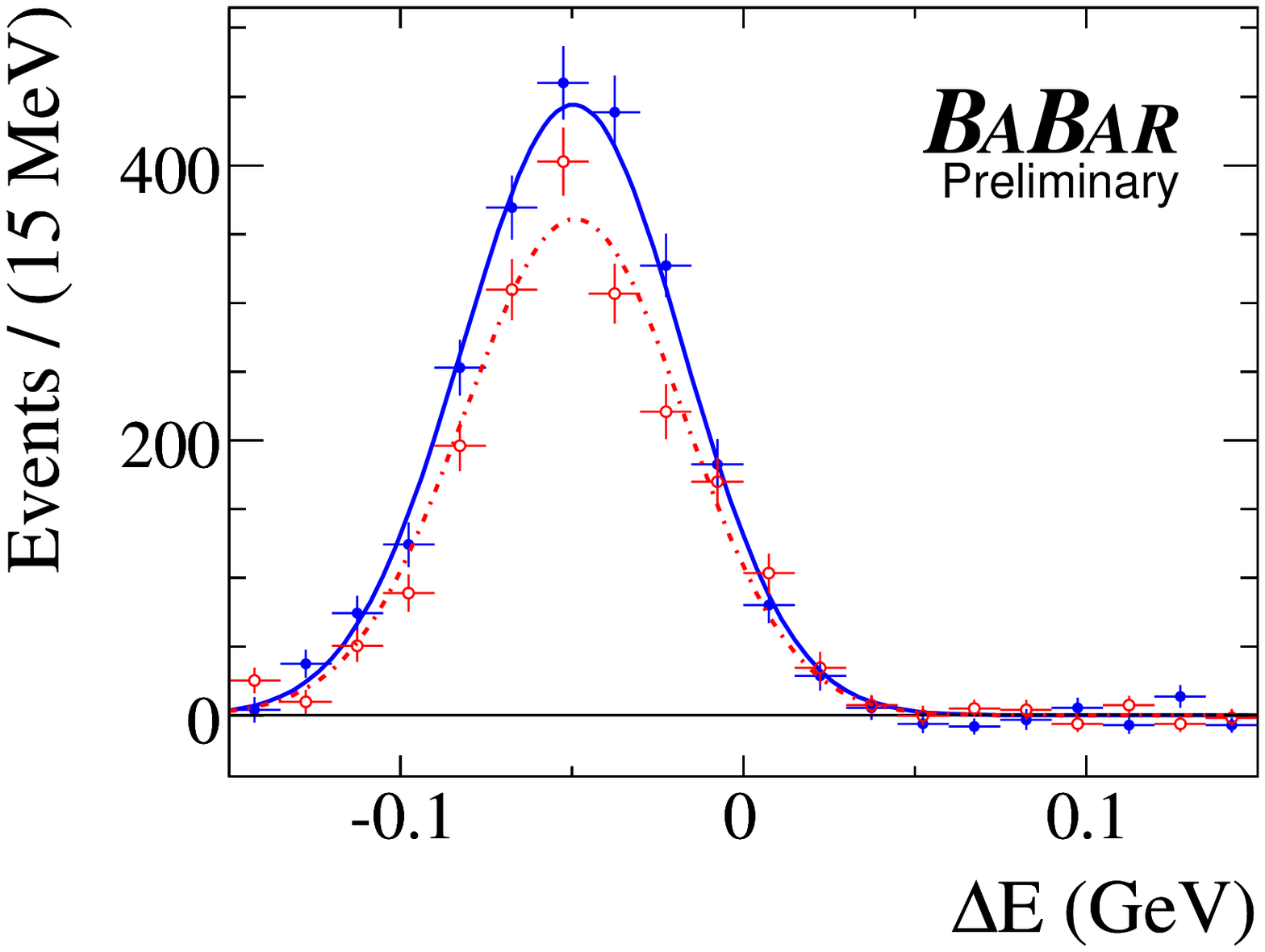,height=1.3in}
\caption{The background-subtracted distribution of $\Delta E$ for signal
$K^{\pm} \pi^{\mp}$ events with data sample 347 M $B \overline{B} (left)$
and 383 M $B \overline{B}$ (right) at BaBar, comparing $B^0$ (solid) and $\overline{B}{}^0$ decays (dashed). }
\label{babar:kpi}
\end{figure}

Throughout this letter, the partial-rate asymmetry is define as $A_{CP}(B \to f) = (\Gamma(\overline{B} \to \overline{f})-\Gamma(B \to f))/(\Gamma(\overline{B} \to \overline{f})+\Gamma(B \to f))$, 
where $\overline{B}$ and $\overline{f}$ are the conjugate states. Belle 
provides the
partial-rate asymmetries for $B^0 \to K^+ \pi^-$, $B^+ \to K^+ \pi^0$ 
and $B^+ \to \pi^+ \pi^0$ three decay modes and the results are shown in 
Fig. \ref{belle:kpi} and Table \ref{tab:acp}.
The partial-rate asymmetry $A_{CP}(K^+ \pi^-)$ is found to be $-0.093 \pm 0.018 \pm 0.008$, which 4.8$\sigma$ from zero. The measurement of $A_{CP}(K^+ \pi^0)$ is consistent 
with no asymmetry; the central value is $4.4\sigma$ away from $A_{CP}(K^+\pi^-)$.
BaBar also provides the partial-rate asymmetry for these three decay 
modes shown in Fig. \ref{babar:kpi} and
Table \ref{tab:acp} and claims the observation of the partial-rate asymmetry 
$-0.107 \pm 0.018 ^{+0.007}_{-0.004}$ with 5.5$\sigma$ \cite{acp}.


\begin{table}[t]
\caption{Summary of branching fractions}
\vspace{0.4cm}
\begin{center}
\begin{tabular}{|c|c|c|c|c|}
\hline
& \mco{2}{|c|}{Belle} & \mco{2}{|c|}{BaBar} \\
\hline
Mode & N$_{\rm signal}$ & BF($10^{-6}$) & N$_{\rm signal}$ & BF($10^{-6}$) \cr
\hline
$K^+ \pi^-$ & 3585$^{+69}_{-68}$ & 19.9 $\pm$ 0.4 $\pm$ 0.8 & 1660$\pm$ 52 & 19.7 $\pm$ 0.6 $\pm$ 0.6 \cr
$\pi^+\pi^-$ & 872 $^{+41}_{-40}$ & 5.1 $\pm$ 0.2 $\pm$ 0.2 & 489 $\pm$ 35 & 5.8 $\pm$ 0.4 $\pm$ 0.3 \cr
$K^+ K^-$ & 2.5 $^{+5.0}_{-3.7}$ & $< 0.41$ & 3 $\pm$ 13 & $< 0.4$ \cr
$K^+ \pi^0$ & 1493$^{+57}_{-55}$ & 12.4 $\pm$ 0.5 $\pm$ 0.6 & 1239 $\pm$ 52& 13.3$\pm$ 0.6 $\pm$ 0.6 \cr
$\pi^+\pi^0$ & 693$^{+46}_{-43}$ & 6.5 $\pm$ 0.4 $^{+0.4}_{-0.5}$ & 572 $\pm$ 53& 5.1 $\pm$ 0.5 $\pm$ 0.3 \cr
$\overline{K}{}^0 K^+$ & 36.6 $^{+9.7}_{-8.3}$ & 1.22 $^{+0.32+0.13}_{-0.28-0.16}$ & 71 $\pm$ 19 & 1.61 $\pm$ 0.44 $\pm$ 0.09 \cr
$K^0 \pi^+$ & 1252 $^{+41}_{-39}$ & 22.8$^{+0.8}_{-0.7}\pm1.3$ & 1072 $\pm$ 46 & 23.9 $\pm$ 1.1 $\pm$ 1.0 \cr
$K^0 \overline{K}{}^0$ & 23.0 $^{+6.5}_{-5.4}$ & 0.87$^{+0.25}_{-0.20}\pm0.09$ & 32 $\pm$8 & 1.08$\pm$ 0.28 $\pm$ 0.11 \cr
$K^0 \pi^0$ & 379$^{+28}_{-27}$ & 9.2 $\pm$ 0.7 $^{+0.6}_{-0.7}$ & 425$\pm$28 & 10.5 $\pm$ 0.7 $\pm$ 0.5 \cr
$\rho^+ K^0$ & - & - & 158$^{+27}_{-26}$ & $8.0^{+1.4}_{-1.3} \pm 0.5$ \cr
\hline 
\end{tabular}
\label{br}
\end{center}
\end{table}
\begin{table}[h]
\caption{Summary of partial-rate asymmetry}
\vspace{0.4cm}
\begin{center}
\begin{tabular}{|c|c|c|c|c|}
\hline
& Belle(535M $B\overline{B}$) & BaBar(347M $B\overline{B}$) & BaBar(383M $B\overline{B}$) \cr
\hline
Mode & $A_{CP}$ & $A_{CP}$ & $A_{CP}$ \cr
\hline
$K^+ \pi^-$ & $-0.093 \pm 0.018 \pm 0.008$ & $-0.108 \pm 0.024 \pm 0.007$ & $-0.107 \pm 0.018 ^{+0.007}_{-0.004}$ \cr
$K^+ \pi^0$ & $0.07 \pm 0.03 \pm 0.01$ & $0.016 \pm 0.041 \pm 0.010$ & - \cr
$\pi^+ \pi^0$ & $0.07 \pm 0.06 \pm 0.01$ & $-0.019 \pm 0.088 \pm 0.014$ & - \cr
\hline 
\end{tabular}
\label{tab:acp}
\end{center}
\end{table}

\section{$B^+ \to \rho^+ K^0$, $B^0 \to \rho^0 \rho^0$ and $B^0 \to \phi K^{*0}$}
The pure penguin $b \to s$ decay process $B^+ \to \rho^+ K^0$ is observed by BaBar 
recently \cite{rhok}. The measured branching fraction is ($8.0^{+1.4}_{-1.3}\pm0.5) \times 10^{-6}$ 
 shown in Table \ref{br} with 7.9$\sigma$ significance and is consistent with 
theoretical prediction with the assumption $p'_V = - p'_P$ \cite{prl46} within the 
uncertainties. The $p'_V$ ($p'_P$) is the amplitude for the spectator quark to appear
in the vector (pseudoscalar) meson. 

The value of $\Delta \alpha$ can be extracted from an analysis of the branching
fractions of the $B$ decays into the full set of isospin-related channels \cite{prl10}. BaBar finds the
evidence for $B^0 \to \rho^0 \rho^0$ with 3.5$\sigma$ significance and 
measures the branching fraction 
$(1.07 \pm 0.33 \pm 0.19) \times 10^{-6}$ \cite{rhorho}. With the 
$B^0 \to \rho^0 \rho^0$ measurement, BaBar obtains a 68\% (90\%) CL limit on 
$|\Delta \alpha| \equiv |\alpha - \alpha_{\rm eff}|<18^o$ $(<20^o)$. An
isospin-triangle relation holds for each of the three helicity amplitudes,
which can be separated through an angular analysis. The longitudinal 
polarization  fraction $f_L = |A_0|^2/(\sum |A_{\lambda}|^2)$ of $\rho^0 \rho^0$
is $0.87 \pm 0.13 \pm 0.04$, where $A_{\lambda=-1,0,+1}$ are the helicity
amplitudes.

The large fraction of tranverse polarization in the $B \to \phi K^*(892)$ decay
measured by Belle \cite{phik1} and BaBar \cite{phik2} indicates a significant
departure from the naive expectation of dominant longitudinal polarization.
BaBar extend their investigation of the polarization puzzle with an amplitude
analysis of the vector-tensor $B^0 \to \phi K^*_2(1430)^0$ decay and 
vector-scalar $B^0 \to \phi (K\pi)^{*0}_0$ decay \cite{phik3}. The amplitudes are
reparameterized with the index $J$ suppressed as $A_0$ and 
$A_{\pm 1}=(A_{\parallel}\pm A_{\perp})/\sqrt{2}$ and the transverse
polarization fraction is defined as $f_{\perp} = |A_{\perp}|^2/\sum|A_{\lambda}|^2$. The polarization results are
\bea
f_L(B^0 \to \phi K^*(892)^0) = 0.506 \pm 0.040 \pm 0.015 \nonumber \\
f_L(B^0 \to \phi K^*_2(1430)^0) = 0.853 ^{+0.061}_{-0.069} \pm 0.036 \nonumber \\
f_{\perp}(B^0 \to \phi K^*(892)^0) = 0.227 \pm 0.038 \pm 0.013 \nonumber \\
f_{\perp}(B^0 \to \phi K^*_2(1430)^0) = 0.045^{+0.049}_{-0.040} \pm 0.013 \nonumber 
\eea
\section{$B \to a_0 K$ and $a_0 \pi$}
BaBar apply separate fits to determine the $a_0(980)$ and $a_0(1450)$ yields 
since this results in $\sim$20\% better sensitivity for $a_0(980)$. The 
$a_0(1450)$ fit has a component for $a_0(980)$ with the yiled fixed to the value
found in the $a_0(980)$ fit, corrected for the small efficiency difference. 
Since the branching fraction for $a_0 \to \eta \pi$ is not well known, the following
is the 90\% C.L. upper limits of product branching fraction \cite{babara0}. Since the
branching fractions of these decays are similar, it means the $a_0(980)$
meson tend to four-quark state \cite{a0}.
\bea
{\cal B}(B^0 \to a_0^- \pi^+) \times (a_0 \to \eta \pi) < 3.1 \times 10^{-6} \nonumber \\
{\cal B}(B^0 \to a_0^- K^+) \times (a_0 \to \eta \pi) < 1.9 \times 10^{-6} \nonumber \\
{\cal B}(B^0 \to a_0(1450)^- \pi^+) \times (a_0 \to \eta \pi) < 2.3 \times 10^{-6} \nonumber \\
{\cal B}(B^0 \to a_0(1450)^- K^+) \times (a_0 \to \eta \pi) < 3.1 \times 10^{-6} \nonumber
\eea

\end{document}